# Task allocation planning based on HTN for national economic mobilization

Peng Zhao

**Abstract** In order to cope with the task allocation in national economic mobilization, a task allocation planning method based on Hierarchical Task Network (HTN) for national economic mobilization is proposed. An HTN planning algorithm is designed to solve and optimize task allocation, and a method is explored to deal with the resource shortage. Finally, based on a real task allocation case in national economic mobilization, an experimental study verifies the effectiveness of the proposed method.

国民经济动员是国家调动经济资源潜力和社会的物力、财力应对紧急事态的活动[1]. 在整个国民经济动员活动中，任务分配是将当前的动员任务合理地分配给企事业单位，以达到快速、经济、可靠地完成动员任务的目的. 动员任务分配的好坏直接决定了动员任务完成的质量，不合理的分配方案有可能造成任务无法完成，导致战争失败或者重大突发事件失控.

王红卫等[2]对国民经济动员任务分配建立多目标模型，将时间、风险因素转化为经济成本对模型进行求解. 郭瑞鹏[3]对国民经济动员任务分配中物资动员决策问题进行建模，求解最优物资动员方案. 然而，传统的国民经济动员任务分配方法只面对固定的理论数学模型，而国民经济动员任务分配问题处于不断变化之中，其无法对变化的问题求解；同时，国民经济动员任务分配要求紧急性与实时性，传统方法对变化的问题重新建立模型需要较长的时间，无法对问题及时响应；并且，传统的理论数学模型缺乏具体行动的执行序列，无法为国民经济动员任务分配提供直接可行的行动方案. 此外，资源缺项广泛存在于国民经济动员任务分配中，如果资源缺项发生，传统方法因不满足约束而无法得到分配方案.

智能规划[4]对若干执行动作和相关资源、时间等进行推理，得出可完成目标任务的动作序列. 对于变化的问题，智能规划推理各执行动作，无需重新修改规划模型；同时，智能规划通过规划算法自动执行，可对动员任务自动快速地响应；并且，其规划得出直接可行的动作序列. 因此，智能规划适用于国民经济动员任务分配，其既可以求解复杂任务约束关系，也可以面向应用得出具体可行的动作序列. 然而，现有的大多数规划方法无法很好地利用领域知识对求解过程进行优化.

层次任务网络 (Hierarchical Task Network, HTN) 规划[5]是一种被广泛应用的智能规划方法[6,7]，其按照任务网络分层分解的方式进行规划推理，将抽象的目标任务作为复合任务，将复合任务不断地递归分解成越来越具体的子任务，直至所有的子任务都为不可分解且可以直接执行的原子任务为止[8,9]. 这与实际国民经济动员任务分配过程非常相似. HTN 规划具有丰富的领域知识表达能力和高效的推理能力[4]，可以处理国民经济动员任务分配这类复杂且规模较大的问题，并且，依靠领域知识与启发式算法，可以得出比较优解甚至最优解. 因此，基于 HTN 的国民经济动员任务分配规划方法是求解动员任务分配问题的有效方法.

本文以国民经济动员任务分配为研究对象，首先对国民经济动员任务分配问题进行分析，提出基于 HTN 的国民经济动员任务分配规划算法，并且设计了一种缺项处理方法，用以解决国民经济动员任务分配过程中的资源缺项问题. 最后，通过实际国民经济动员任务分

配案例，验证该方法的有效性.

# 1. 任务分配规划问题描述

国民经济动员任务分配就是在不同时间、资源等约束下，完成对动员物资的生产并运输至集结地的目标. 具体的目标为: 一是在任务目标时间前生产动员物资并运输至集结地; 二是任务总成本尽可能小.

**定义1** 国民经济动员任务分配规划问题可描述为 $P = (s_0, \Sigma, g)$，其中 $s_0$ 为初始状态，描述了规划初始时刻的状态; $\Sigma$ 为规划领域，指完成规划的领域知识; $g$ 为目标任务，指目标任务集合. 规划领域可以描述为 $\Sigma = (O, M)$，其中 $O$ 为操作集合，$M$ 为方法集合. 操作与方法的定义与经典 HTN 规划[10,11,12]一致.

**定义 2** 国民经济动员任务分配规划目标任务 $g = \{T_1, T_2, \ldots, T_n\}$ 即上级单位下达的任务集合，其中 $T = \{TaskID, TaskDeadline, TaskAmount, TaskProductID, TaskDestination\}$, $T$ 是国民经济动员目标任务, $TaskID$ 是国民经济动员任务编号, $TaskDeadline$ 是完成生产和运输的总时间, $TaskProductID$ 是完成生产和运输的动员物资编号, $TaskAmoun$ 是生产和运输动员物资的数量, $TaskDestination$ 是动员物资集结地.

**定义3** 规划解 $\pi$ 为规划问题的解，表示为 $\pi = \{A_1, A_2, \ldots, A_n, R_1, R_2, \ldots, R_n\}$，其中 $A$ 为动作，$R$ 为缺项报告. $\pi$ 即为求解规划问题 $P$ 的方案.

# 2. 任务分配规划

## 2.1 基于 HTN 的任务分配规划算法

基于 HTN 的国民经济动员任务分配规划方法以问题知识和领域知识作为输入，输出规划解包括动作序列和缺项报告，其总体框架如图 1 所示.

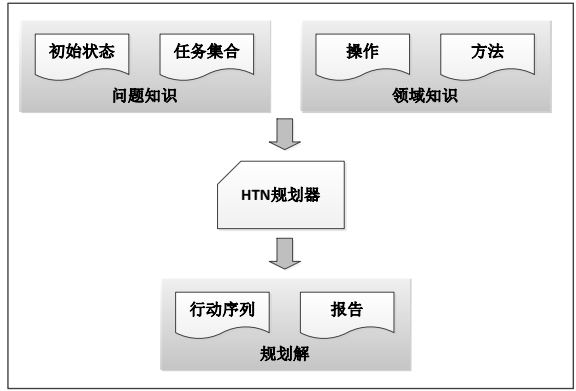

图 1 基于 HTN 的国民经济动员任务分配规划总体框架

在基于 HTN 的国民经济动员任务分配规划中，复合任务在满足各种约束下被方法分解为更加具体的子任务，直至可直接执行的动作，即得到了规划解. 具体规划算法如图 2 所示.

| 规划算法: $Plan(s, \Sigma, g)$ |
|---|
| Initialize: $s \leftarrow s_0$ |
| 1. **if** $g \neq \varnothing$ **then** |
| 2. pop a $T \in g$ by max $\gamma(TaskDeadline, TaskAmount)$ |
| 3. **if** $T$ is a primitive task **then** |
| 4. $Active \leftarrow \{a \mid a$ is a ground instance of an operator in $O$, $a$ is applicable to $s\}$ |
| 5. **while** $Active \neq \varnothing$ **do** |
| 6. non-deterministically pop $a \in Active$ |
| 7. **if** ResourceShortage $m$ occurs in $T$ **then** |
| 8. $R \leftarrow R \bigcup m$ |
| 9. disposal the ResourceShortage |

| 10. | $s' \leftarrow (s - effect^-(a)) \bigcup effect^+(a)$ |
|---|---|
| 11. | $g' \leftarrow g - T$ |
| 12. | $\pi \leftarrow Plan(s', \Sigma, g') \bigcup a$ |
| 13. | **if** $\pi$ is not failure **then return** $\pi$ |
| 14. | **end while** |
| 15. | **if** $T$ cannot be completed before *TaskDeadline* **then** |
| 16. | $R \leftarrow R \bigcup T$ |
| 17. | **if** $Active = \varnothing$ **then return** failure |
| 18. | **else if** $T$ is a compound task **then** |
| 19. | $Active \leftarrow \{ m \mid m$ is a ground instance of a method in $M$, $m$ is applicable to $s \}$ |
| 20. | **while** $Active \neq \varnothing$ **do** |
| 21. | pop $m \in Active$ by $\max \gamma(production, cost)$ or $\max \gamma(speed, cost)$ |
| 22. | $g' \leftarrow g - T \bigcup subtasks$ |
| 23. | $\pi \leftarrow Plan(s, \Sigma, g')$ |
| 24. | **if** $\pi$ is not failure **then return** $\pi$ |
| 25. | **end while** |
| 26. | **if** $T$ cannot be completed before *TaskDeadline* **then** |
| 27. | $R \leftarrow R \bigcup T$ |
| 28. | **if** $Active = \varnothing$ **then return** failure |
| 29. | **else if** $g = \varnothing$ **then return** $\pi$ |

图 2 基于 HTN 的任务分配规划算法

第 2 行根据任务效能函数选择当前任务. 任务效能函数 $\gamma$ (*TaskDeadline, TaskAmount*) = *TaskAmount / TaskDeadline*, 表示任务的紧急程度, $\gamma$ 越大, 表示任务越紧急, 要求集中时间与资源优先满足. 优先规划紧急任务可以使时间与资源得到合理分配, 最大限度地保证了应急需求.

当前任务为原子任务时（第 3-17 行）, 对操作实例化并选择一个动作实例, 被选择动作改变状态. 如果资源不足, 则上报资源缺项并处理 （第 7-9 行）. 缺项处理具体算法将在 2.2 节讨论.

当前任务为复合任务时（第 18-28 行）, 对方法实例化并选择子任务实例, 替换当前任务. 其中, 当复合任务选择开启生产线时, 其根据生产线效能函数选择子任务实例. 生产线效能函数 $\gamma$ (*production, cost*) = *production / cost*, 表示生产线生产能力成本比, $\gamma$ 越大, 表示生产线效率越高. 优先选择开启生产能力成本比高的生产线在保证完成任务的前提下降低生产成本. 当复合任务选择运输车辆时, 其根据车辆效能函数选择子任务实例. 车辆效能函数 $\gamma$ (*speed, cost*) = *speed / cost*, 表示车辆速度成本比. 优先选择速度成本比高的车辆执行运输任务.

## 2.2 缺项处理

缺项是当企业分析现有资源库存不足以完成生产任务时, 对相应数量的缺项资源向上级单位上报, 请求调拨相应的资源以完成生产任务. 由于国民经济动员任务的不确定且资源消耗较大, 资源缺项广泛存在于国民经济动员任务中, 缺项的产生将直接导致相关任务无法实现. 因此, 根据国民经济动员的实际需求, 基于 HTN 的国民经济动员任务分配规划产生带有缺项报告的执行动作序列, 增强了规划解的实用性和鲁棒性. 具体来说, 国民经济动员任务分配规划首先分析目标任务和相应资源, 如果存在资源缺项, 则上报上级单位; 同时, 为保证及时响应动员任务和产生完整任务分配方案, 虚拟出相应的缺项物资来满足规划的约束条件. 当上级单位调拨缺项资源满足需求后, 其规划解即可立即执行.

缺项识别操作描述为 O = (ResourceShortage taskID mtrID lackamount), 表示为完成编号 TaskID 任务, 企业现有库存缺少编号 mtrID 原材料 lackamount 单位. 故企业向上级单位上报相应原材料的缺失数量, 以保证原材料充分.

缺项处理算法如图 3 所示.

| | |
|---|---|
| 1. | **loop** |
| 2. | analyze mobilized material $M = \{M_1, M_2, \ldots, M_n\}$ |
| 3. | stored material $SM = \{SM_1, SM_2, \ldots, SM_n\}$ |
| 4. | **If** $M_x > SM_x$ **then** |
| 5. | report $M_x$ |
| 6. | handle $M_x$, assume $M_x - SM_x$ |
| 7. | **end loop** |

图 3 缺项处理算法

算法分析当前任务使用的每一种资源. 如果某资源使用量超出其库存量（第 4 行），则上报该资源缺项，并虚拟出相应资源保证规划进行（第 5-6 行）.

## 3. 案例与分析

本文以国家国民经济动员现实情况为案例，利用基于 HTN 的国民经济动员任务分配规划方法，得出国民经济动员任务分配的方案. 本文案例实现基于 HTN 规划器 SHOP2[13]，运行于配置 3.1 GHz CPU, 4GB 内存的 Windows 7 操作系统.

### 3.1 剧情介绍

企业 a1 接受上级单位下达的任务，该企业具备生产和运输能力，对原材料有一定库存. 以下是对剧情相应信息的介绍，为简单起见，忽略以下描述中数量的单位. 企业资源总量如表 1 所示. 原材料信息如表 2 所示. 生产线生产能力及成本如表 3 所示. 生产线资源消耗期刊如表 4 所示. 运输车辆信息如表 5 所示. 运输距离如表 6 所示. 装载、卸载速度如表 7 所示.

表 1 企业资源总量

| | 水 | 电 | 汽 | 工人 |
|---|---|---|---|---|
| 资源总量 | 80000 | 80000 | 80000 | 1000 |

表 2 原材料信息

| 原材料库存 | | m001 | m002 | m003 | m004 | m005 | m006 |
|---|---|---|---|---|---|---|---|
| | | 10000 | 10000 | 10000 | 10000 | 10000 | 10000 |
| 动员物资所需原材料 | p001 | 2 | 3 | 5 | - | - | - |
| | p002 | 1 | 2 | - | 3 | - | - |
| | p003 | - | - | 5 | - | 3 | 1 |

表 3 生产线生产能力及成本

| | | p001 | p002 | p003 |
|---|---|---|---|---|
| 生产能力 | l001 | 20 | 25 | 30 |
| | l002 | - | 40 | - |
| | l003 | 30 | - | - |
| | | p001 | p002 | p003 |
| 成本 | l001 | 10 | 20 | 20 |
| | l002 | - | 50 | - |
| | l003 | 40 | - | - |

表 4  生产线资源消耗

| 生产线 | l001 | | | l002 | | | l003 | | |
|---|---|---|---|---|---|---|---|---|---|
| 动员物资 | p001 | p002 | p003 | p001 | p002 | p003 | p001 | p002 | p003 |
| 水 | 50 | 40 | 20 | - | 90 | - | 50 | - | - |
| 电 | 60 | 30 | 30 | - | 60 | - | 40 | - | - |
| 汽 | 60 | 50 | 50 | - | 60 | - | 40 | - | - |
| 工人 | 30 | 30 | 40 | - | 30 | - | 20 | - | - |

表 5  运输车辆信息

| 速度 | | c001 | c002 | c003 | c004 | c005 | c006 | c007 | c008 |
|---|---|---|---|---|---|---|---|---|---|
| | | 70 | 90 | 70 | 90 | 90 | 70 | 70 | 70 |
| | | c001 | c002 | c003 | c004 | c005 | c006 | c007 | c008 |
| 运输能力 | p001 | 60 | 50 | 60 | 20 | 20 | - | - | - |
| | p002 | 60 | 50 | - | - | 60 | 50 | 50 | - |
| | p003 | 60 | 50 | - | - | - | - | 60 | 50 |

表 6  企业所在地与集结地路程

| | b1 | b2 | b3 |
|---|---|---|---|
| a1 | 100 | 120 | 80 |

表 7  装载、卸载速度

| | p001 | p002 | p003 |
|---|---|---|---|
| 装载速度 | 50 | 60 | 50 |
| 卸载速度 | 50 | 50 | 60 |

### 3.2 优化规划解案例

目标任务 g= Task1，其中 Task1={t001, 9, 200, p001, b1}，即编号 t001 任务 Task1 在 9 小时内生产并运输 200 件编号 p001 动员物资到集结地 b1.

| Task1 | Task2, Task3 |
|---|---|
| [1]  (!start l003 0.0 t001) | [1]  (!start l003 0.0 t002) |
| [2]  (!start l001 0.0 t001) | [2]  (!start l001 0.0 t002) |
| [3]  (!load c001 t001 p001 60.0 1.2) | [3]  (!load c001 t002 p001 60.0 1.2) |
| [4]  (!transport c001 t001 p001 60.0 2.4) | [4]  (!transport c001 t002 p001 60.0 2.4) |
| [5]  (!unload c001 t001 p001 60.0 3.8) | [5]  (!unload c001 t002 p001 60.0 3.8) |
| [6]  (!back c001 t001 p001 5.0) | [6]  (!back c001 t002 p001 5.0) |
| [7]  (!load c003 t001 p001 60.0 2.4) | [7]  (!load c003 t002 p001 40.0 2.0) |
| [8]  (!transport c003 t001 p001 60.0 3.6) | [8]  (!transport c003 t002 p001 40.0 2.8) |
| [9]  (!unload c003 t001 p001 60.0 5.0) | [9]  (!unload c003 t002 p001 40.0 4.2) |
| [10]  (!back c003 t001 p001 6.2) | [10]  (!back c003 t002 p001 5.0) |
| [11]  (!load c002 t001 p001 50.0 3.4) | [11]  (ResourceShortage t003 m001 100.0) |
| [12]  (!transport c002 t001 p001 50.0 4.4) | [12]  (!start l002 0.0 t003) |
| [13]  (!unload c002 t001 p001 50.0 5.5) | [13]  (!load c006 t003 p002 50.0 1.2) |
| [14]  (!back c002 t001 p001 6.5) | [14]  (!transport c006 t003 p002 50.0 2.1) |
| [15]  (!load c001 t001 p001 30.0 6.5) | [15]  (!unload c006 t003 p002 50.0 3.5) |
| [16] (!transport c001 t001 p001 30.0 7.1) | [16] (!back c006 t003 p002 4.5) |
| [17] (!unload c001 t001 p001 30.0 8.5) | [17] (!load c006 t003 p002 50.0 5.9) |
| [18] (!back c001 t001 p001 9.1) | [18] (!transport c006 t003 p002 50.0 6.8) |
| | [19] (!unload c006 t003 p002 50.0 8.2) |
| | [20] (!back c006 t003 p002 9.2) |
| | [21] (!start l001 2.5 t003) |
| | [22] (!load c006 t003 p002 50.0 10.6) |
| | [23] (!transport c006 t003 p002 50.0 11.5) |
| | [24] (!unload c006 t003 p002 50.0 12.9) |
| | [25] (!back c006 t003 p002 13.9) |

图 4 规划结果

规划结果如图 4 所示, 在 0 时刻, 企业开启生产线 l003 与 l001 开始生产任务 t001 动员物资 p001. 在 1.2 时刻, 选择车辆 c001 运输 60 件动员物资 p001; 在 2.4 时刻, 选择车辆 c003 运输 60 件动员物资 p001; 在 3.4 时刻, 选择车辆 c002 运输 50 件动员物资 p001; 在 6.5 时刻, 选择车辆 c001 运输 30 件动员物资 p001.

由于任务目标时间充足, 运输任务中选择了成本较低但行驶速度较慢的车型运输, 在目标时间内完成任务的前提条件下, 得到了成本较低的运输方案, 即规划找到可行解的同时, 找到比较优的可行解.

仔细分析此案例的规划解可知, 在企业现有的情况下, 在此任务中得到的规划解是当前条件下最优解, 即满足任务要求的成本最低生产运输方案. 由此方案显示, 在某些条件下在规划算法的引导下可得到最优解.

### 3.3 缺项处理案例

目标任务 g = {Task2, Task3}, 其中 Task2={t002, 7, 100, p001,b1}, Task3={t003, 20, 150, p002, b1}, 即编号 t002 任务 Task2 在 7 小时内生产并运输 100 件编号 p001 动员物资到集结地 b1, 编号 t003 任务 Task3 在 20 小时内生产并运输 150 件编号 p002 动员物资到集结地 b1. 初始原材料库存如表 8 所示, 其他环境信息与 3.1 节一致.

表 8 原材料库存

| m001 | m002 | m003 | m004 | m005 | m006 |
|---|---|---|---|---|---|
| 250 | 1000 | 1000 | 1000 | 1000 | 1000 |

规划结果如图 4 所示, 规划解分别实现任务 Task2 (第 1-10 行) 和任务 Task3 (第 11-25 行). 规划模型分析原材料, 任务 t003 对原材料 m001 缺少 100 件, 相应的缺项向上级单位上报 (第 11 行).

在 0 时刻, 企业开启生产线 l003 与 l001 开始生产任务 t002 动员物资 p001. 在 1.2 时刻, 选择车辆 c001 运输 60 件动员物资 p001; 在 2.0 时刻, 选择车辆 c003 运输 40 件动员物资 p001.

在 0 时刻, 开启生产线 l002 开始生产任务 t003 动员物资 p002. 在 1.2 时刻, 选择车辆 c006 运输 50 件动员物资 p002. 在 2.5 时刻, 生产线 l001 完成了任务 t002 的生产, 转换生产任务 t003 动员物资 p002. 在 5.9 时刻, 选择车辆 c006 运输 50 件动员物资 p002, 在 10.6 时刻, 选择撤离 c006 运输 50 件动员物资 p002.

在本次案例中, 任务 Task3 的资源缺项是由于任务 Task2 生产消耗引起的. 如果任务 Task2 取消或上级单位调拨了相应缺项的原材料, 则任务 Task3 可立即按规划解执行.

规划结果显示, 多任务间的耦合使任务 Task2 对原材料的消耗造成任务 Task3 产生缺项, 通过本文所述缺项处理方法可在多任务下动态应对资源缺项.

## 4. 结论

本文提出了一种基于 HTN 的国民经济动员任务分配规划方法，求解并优化动员任务分配问题，同时提出了处理物资缺项的方法. 最后，以国民经济动员任务分配中实际案例验证所研究方法的有效性. 通过分析案例结果显示，该方法能够有效规划得到任务分配方案并应对资源缺项的发生.

## 参考文献

[1] 陈德第. 新时期国民经济动员理论框架[J]. 北京理工大学学报：社会科学版，2003, 5(3):44-48.

[2] 王红卫，洪丽琼，陈曦. 国民经济动员中的任务分配问题研究[J]. 东南大学学报(哲学社会科学版), 2009, 11(1):56-59.

[3] 郭瑞鹏. 应急物资动员决策的方法与模型研究[D]. 北京：北京理工大学, 2006.

[4] Ghallab M, Nau D, Traverso P. Automated planning: theory & practice[M]: Elsevier, 2004.

[5] Qi C, Wang D, Munoz-Avila H, Zhao P and Wang H. Hierarchical task network planning with resources and temporal constraints[J]. Knowledge-Based Systems, 2017, 113:17–32.

[6] Liu D, Wang H, Qi C, Zhao P, Wang J. Hierarchical task network-based emergency task planning with incomplete information, concurrency and uncertain duration[J]. Knowledge-Based Systems, 2016, 112:67-79.

[7] Hayashi H, Tokura S, Ozaki F, Hasegawa T. Emergency HTN Planning. Intelligent Automation and Computer Engineering[M], Springer, 2010: 27-40.

[8] Zhao P, Wang H, Qi C and Liu D. HTN planning with uncontrollable durations for emergency decision-making[J]. Journal of Intelligent & Fuzzy Systems, 2017, 33:255–267.

[9] Wang H, Liu D, Zhao P, C. Qi and Chen X. Review on hierarchical task network planning under uncertainty[J]. Acta Automatica Sinica 2016, 42(5):655-667.

[10] Georgievski I, Aiello M. HTN planning: Overview, comparison, and beyond[J]. Artificial Intelligence, 2015, 222:124-156.

[11] Zhao P. Probabilistic contingent planning based on HTN for high-quality plans[J]. arXiv preprint arXiv:2308.06922, 2023.

[12] Zhao P, Qi C and Liu D. Resource-constrained Hierarchical Task Network planning under uncontrollable durations for emergency decision-making[J]. Journal of Intelligent & Fuzzy Systems, 2017, 33:3819-3834.

[13] Nau D, Au T-C, Ilghami O, Kuter U, Murdock JW, Wu D, Yaman F. SHOP2: An HTN Planning System[J]. Journal of Artificial Intelligence Research, 2003, 20:379-404.